\documentclass[aps,prd,twocolumn,superscriptaddress,nofootinbib,floatfix]{revtex4-2}
\usepackage{amsmath,amssymb}
\usepackage{graphicx}
\usepackage{bm}
\usepackage{booktabs}
\usepackage[colorlinks=true,linkcolor=blue,citecolor=blue,urlcolor=blue]{hyperref}

\newcommand{\threebar}{\bar{\bm 3}}
\newcommand{\Sym}{\mathrm{Sym}}

\begin{document}

\title{Diquark matter in the color-spin model:\\ color--flavor correspondence and Pauli blocking}

\author{Aaron Park}
\email{aaron.park@yonsei.ac.kr}
\affiliation{Department of Physics and Institute of Physics and Applied Physics, Yonsei University, Seoul 03722, Korea}

\date{\today}

\begin{abstract}
We study the color-spin interaction of a quark, a scalar diquark, an
$H$-dibaryon-like six-quark state, and an octet baryon in a medium of
scalar diquarks sharing a common $s$-wave orbital. Constructing the medium
with diquark creation operators yields matching color and flavor
representations and excludes many otherwise Pauli-allowed quark states.
For all four probes, the trace-averaged interaction per surrounding
diquark is independent of the diquark number $n$ for $n\le3$. Additional
Pauli constraints break this plateau at $n=4$. Including the internal
color-spin energy, the six-quark probe has the lowest addition cost per
quark for $n\le3$, while a single quark has the lowest cost at most higher
fillings. In a number-projected color-flavor-locked condensate, a coherent
pair of the condensate's own type minimizes the addition cost at every
filling studied. The color-spin interaction favors scalar pairing and
provides no additional binding between diquarks in the common orbital.
\end{abstract}

\maketitle

\section{Introduction}

Recent observations of massive neutron stars~\cite{Demorest:2010bx,Antoniadis:2013pzd,NANOGrav:2019jur}, gravitational waves from neutron star mergers~\cite{LIGOScientific:2017vwq,LIGOScientific:2018cki}, and NICER radius measurements~\cite{Miller:2021qha,Riley:2021pdl} require an equation of state (EOS) that stiffens rapidly above nuclear saturation density.
Quarkyonic matter~\cite{McLerran:2007qj}, in which a quark Fermi sea is surrounded by a baryonic shell in momentum space, provides a natural mechanism for such stiffening~\cite{McLerran:2018hbz,Jeong:2019lhv,Duarte:2020xsp,Duarte:2020kvi}.
At high density, quark matter is expected to be a color superconductor~\cite{Alford:1997zt,Rapp:1997zu,Alford:1998mk,Alford:2007xm}, in which quarks form Cooper pairs in the most attractive scalar, color-antitriplet, flavor-antitriplet channel~\cite{Jaffe:2004ph,Anselmino:1992vg}.
Between the dilute and the asymptotic regimes, diquarks may appear as tightly bound bosons. Their Bose--Einstein condensation and its crossover to BCS pairing have been studied in Nambu--Jona-Lasinio models~\cite{Sun:2007fc,Abuki:2010jq}, and the analogous regime of two-color QCD is accessible to lattice simulations~\cite{Braguta:2016cpw,Iida:2022hyy}.
In these approaches the diquark is a field or a point boson, and the Pauli principle among its constituent quarks enters only implicitly.
The extent to which a dense collection of diquarks can be treated as bosons at all has also been questioned for light-quark multiquark states~\cite{Gal:2019xju}.

In a series of works, we have studied the short-distance part of this problem in a constituent quark model, where the color-spin interaction controls the repulsive core between hadrons~\cite{Park:2019bsz}.
By comparing the color-spin interaction felt by a quark and a baryon in a baryonic medium, we found a quarkyoniclike configuration at high density~\cite{Park:2021hqb}, identified the most stable diquark in dense matter~\cite{Park:2022qew}, and argued for a diquarkyonic configuration in which quarks and diquarks coexist~\cite{Park:2022jpf}.
The natural next question, raised in Ref.~\cite{Park:2022jpf}, is what happens when the medium itself is made of diquarks, that is, how a quark, a diquark, or a baryon interacts with a dense ensemble of diquarks and whether the diquarks are stable there.
Answering it requires keeping the quark substructure of every diquark explicitly, which is the aim of this work.

In this work we address this question with two methodological improvements.
First, we construct the multiquark state space explicitly as the span
of the states obtained by applying diquark creation operators to the
probe state. Because these operators are built from quark operators,
the Pauli principle is enforced automatically.
For bosonic diquarks the Cauchy identity makes the color and flavor representations of the medium coincide, a property we call the color--flavor correspondence. Many color--flavor--spin states that are compatible with the Pauli principle at the quark level, and that would be retained if one used only the outer-product selection rules, are therefore absent.
Second, besides the state-averaged interaction used in our previous works, we consider (i) the lowest state, (ii) the sector in which the total system is a color singlet, where the
result turns out to be unique, and (iii) number-projected 2SC and CFL condensates.

We find a simple structural result. As long as the diquarks behave as elementary bosons, the averaged interaction per surrounding diquark does not depend on the number of diquarks. The departure of this averaged interaction from its plateau therefore comes entirely from the quark substructure of the diquarks, which limits how many of them fit into one orbital (Sec.~\ref{sec:theorem}).
Including the internal interaction of the probes, an equally weighted medium favors the three correlated diquarks for $n\le3$ and a single quark at most higher fillings. A CFL condensate favors further pairing of its own type, and in the 2SC medium the addition energies favor pairing channels compatible with CFL order.

The paper is organized as follows.
In Sec.~\ref{sec:model}, we define the interaction.
Section~\ref{sec:states} constructs the state space and Sec.~\ref{sec:theorem} derives the $n$ independence.
Results are presented in Sec.~\ref{sec:results}, their robustness is examined in Sec.~\ref{sec:robust}, and Sec.~\ref{sec:summary} summarizes.
The numerical method is given in Appendix~\ref{app:method}, and the calculation in the $0s+0p$ space in Appendix~\ref{app:pwave}.

\section{Color-spin interaction}
\label{sec:model}

We assume that all quarks occupy the same spatial orbital, so that the spatial part of the wave function is totally symmetric and the strength of each pairwise interaction is determined by its color and spin factors only~\cite{Park:2019bsz,Park:2021hqb}.
The color-spin interaction is
\begin{align}
V_{CS} = -\sum_{i<j}\frac{1}{m_i m_j}\lambda^c_i\lambda^c_j\,\bm\sigma_i\cdot\bm\sigma_j \equiv \frac{1}{m_u^2}H_{CS},
\label{eq:vcs}
\end{align}
where $\lambda^c_i$ are the Gell-Mann matrices for color SU(3) and $m_u$ is the constituent mass of the light quarks.
In the flavor SU(3) symmetric limit, for $N$ quarks in a totally antisymmetric color$\otimes$flavor$\otimes$spin state,
\begin{align}
H_{CS} &= N(N-10)+\frac{4}{3}S(S+1)+4C_F+2C_C, \label{eq:hcs3}\\
C_R &= \frac{1}{3}\left(p^2+q^2+3p+3q+pq\right), \nonumber
\end{align}
where $(p,q)$ are the Dynkin labels of the flavor ($C_F$) or color ($C_C$) representation.
For flavor SU(2),
\begin{align}
H_{CS} = \frac{4}{3}N(N-6)+\frac{4}{3}S(S+1)+4I(I+1)+2C_C .
\label{eq:hcs2}
\end{align}
For reference, $H_{CS}=-8$ for the scalar diquark $(\threebar_c,\threebar_f,S=0)$ and the octet baryon, and $-24$ for the flavor-singlet, color-singlet, spin-zero six-quark state with the quantum numbers of the $H$ dibaryon~\cite{Jaffe:1976yi}, which we call three correlated diquarks~\cite{Park:2022jpf}.

In terms of the operators $c^\dagger_\alpha$ that create a quark in the
single-particle state $\alpha=(a,i,s)$ of color $a$, flavor $i$, and spin $s$
in the common orbital,
\begin{align}
H_{CS} &= -\frac{1}{2}\sum_{a=1}^{8}\sum_{k=1}^{3} Q^{ak}Q^{ak} + 8N,\nonumber\\
Q^{ak}&=\sum_{\alpha\beta} c^\dagger_\alpha\left(\lambda^a\sigma^k\right)_{\alpha\beta}c_\beta,
\label{eq:hsecond}
\end{align}
which we use to compute expectation values directly in states that are not
decomposed into multiplets (Sec.~\ref{sec:condensate}) and for unequal quark
masses (Sec.~\ref{sec:robust}).

To measure how a probe $p$ interacts with $n$ surrounding diquarks we use, as in Refs.~\cite{Park:2021hqb,Park:2022jpf},
\begin{align}
\Delta H^{\rm avg}_{CS} = \frac{1}{n_p\,n}\left[\langle H_{CS}\rangle_{nd+p}-\langle H_{CS}\rangle_{nd}-H^{p}_{CS}\right],
\label{eq:dH}
\end{align}
where $n_p$ is the number of quarks in the probe, $\langle\cdot\rangle_X$ denotes the trace average over the physical state space of system $X$ constructed in Sec.~\ref{sec:states}, and $H^p_{CS}$ is the internal factor of the probe.
The trace average is equivalent to the weighted average with weights $d_Cd_Fd_S m$ used in Refs.~\cite{Park:2021hqb,Park:2022jpf}, provided the multiplicities $m$ are those of the physical space.
Dividing by $n$ compares different $n$ at fixed density of the medium.
To compare probes that differ in their own internal interaction we also need the total cost of adding the probe. The energy per added quark is
\begin{align}
\mu_p=\frac{\langle H_{CS}\rangle_{nd+p}-\langle H_{CS}\rangle_{nd}}{n_p}=n\,\Delta H_{CS}+\frac{H^p_{CS}}{n_p}.
\label{eq:mu}
\end{align}
The factors $\Delta H_{CS}$ and $\mu_p$ are dimensionless and become energies after multiplication by the spatial coefficient $\kappa$ of the color-spin interaction, which depends on the size $b$ of the common orbital.
If $b$ is kept fixed, the density grows as $2n/b^3$, $\kappa$ is constant, and the color-spin energy per added quark is $\kappa\mu_p$.
To compare different $n$ at the same density instead, the orbital must grow as $b^3\propto n$. For a contact interaction $\kappa\propto b^{-3}$ then scales as $1/n$, $\kappa(n)=\kappa_*/n$ with $\kappa_*$ the value at $n=1$, and the energy per added quark is $\kappa(n)\mu_p=\kappa_*\epsilon_p$ with
\begin{align}
\epsilon_p=\frac{\mu_p}{n}=\Delta H_{CS}+\frac{H^p_{CS}}{n\,n_p},
\label{eq:eps}
\end{align}
which we quote alongside $\Delta H_{CS}$. At fixed $n$, $\epsilon_p$ and $\mu_p$ order the probes in the same way.
Because all $2n+n_p$ quarks are antisymmetrized in one orbital, the probe is correlated with every surrounding diquark, and its internal pairs and its pairs with the medium have the same spatial strength. The internal term therefore scales with the orbital size in the same way as the probe--medium term, and the factor $1/n$ follows from the fixed-density condition $b^3\propto n$.
This scaling holds for the contact color-spin term only. The Coulomb, linear, and constant color-electric terms scale as $b^{-1}$, $b$, and $b^0$ (Sec.~\ref{sec:robust}).

\section{State space of diquark matter}
\label{sec:states}

\subsection{Diquark operators and the Cauchy identity}

Let $c^\dagger_{a i s}$ create a quark with color $a$, flavor $i$, and spin $s$ in the common orbital. There are $3\times3\times2=18$ single-particle states.
The scalar good diquark is
\begin{align}
D^\dagger_{\bar a\bar i} = \sum_{bc}\sum_{jk}\epsilon_{abc}\,\epsilon_{ijk}\,c^\dagger_{bj\uparrow}c^\dagger_{ck\downarrow},
\label{eq:diquark}
\end{align}
which spans the nine-dimensional space $V=(\threebar_c,\threebar_f,S=0)$.
The barred indices indicate this antitriplet character. The $D^\dagger$'s commute with each other, so the states of $n$ diquarks span the image of the symmetric power $\Sym^n(V)$ in the $2n$-quark Fock space.
By the Cauchy identity~\cite{Macdonald:1995,Howe:1989},
\begin{align}
\Sym^n(\threebar_c\otimes\threebar_f)=\bigoplus_{\lambda\vdash n,\ \ell(\lambda)\le 3} S_\lambda(\threebar)_c\otimes S_\lambda(\threebar)_f ,
\label{eq:cauchy}
\end{align}
with total spin $S=0$. In other words, the color and flavor representations of the diquark medium coincide and each pair appears once. We call this the color--flavor correspondence. It holds for a medium built only from scalar diquarks and is
absent for general $2n$-quark states. It is a property of the state space, valid for every state built from scalar diquarks, and should be distinguished from color--flavor locking in the CFL phase, which refers to a particular condensate (Sec.~\ref{sec:condensate}).
For $n=2$ this gives $(\bar{\bm6},\bar{\bm6})\oplus(\bm 3,\bm 3)$, which expresses that a symmetric two-boson state is either symmetric in both color and flavor or antisymmetric in both. For $n=3$ it gives $(\overline{\bm{10}},\overline{\bm{10}})\oplus(\bm 8,\bm 8)\oplus(\bm 1,\bm 1)$.

A state that violates the correspondence, e.g., an eight-quark state with color $\bm 3$ and flavor $\overline{\bm{15}}$, can satisfy the Pauli principle at the quark level and appear in both $(\threebar)^4$ products. It cannot, however, be built from four good diquarks.
Selecting states only by the outer products $(\threebar)^n\otimes(\cdot)$ in color and flavor separately therefore overcounts the state space for $n\ge2$.

\subsection{Pauli blocking}

For $n\le4$, distinct boson states remain linearly independent when written
in terms of quarks, i.e., the map $\Sym^n(V)\to\Lambda^{2n}(\mathbb C^{18})$
that sends a boson state to $D^\dagger\cdots D^\dagger|0\rangle$ is injective,
and the dimensions of the $n$-diquark spaces are $\binom{n+8}{8}=9,45,165,495$.
For $n\ge5$, not every boson state can be realized with quarks, because the
18 single-quark states do not suffice and some products
$D^\dagger\cdots D^\dagger|0\rangle$ vanish or become linearly dependent.
We refer to this reduction as quark-level Pauli blocking.
Because the map commutes with the color and flavor transformations and each $\lambda$ appears only once in Eq.~\eqref{eq:cauchy}, Pauli blocking can only remove entire components $\lambda$. The surviving states therefore keep the color--flavor correspondence.
Table~\ref{tab:ndiquark} lists the content of the $n$-diquark space obtained by the method of Appendix~\ref{app:method}.
The largest number of diquarks is nine, which fills all 18 orbitals, and the resulting state is the unique color-singlet, flavor-singlet state.

\begin{figure*}[t]
\includegraphics[width=\textwidth]{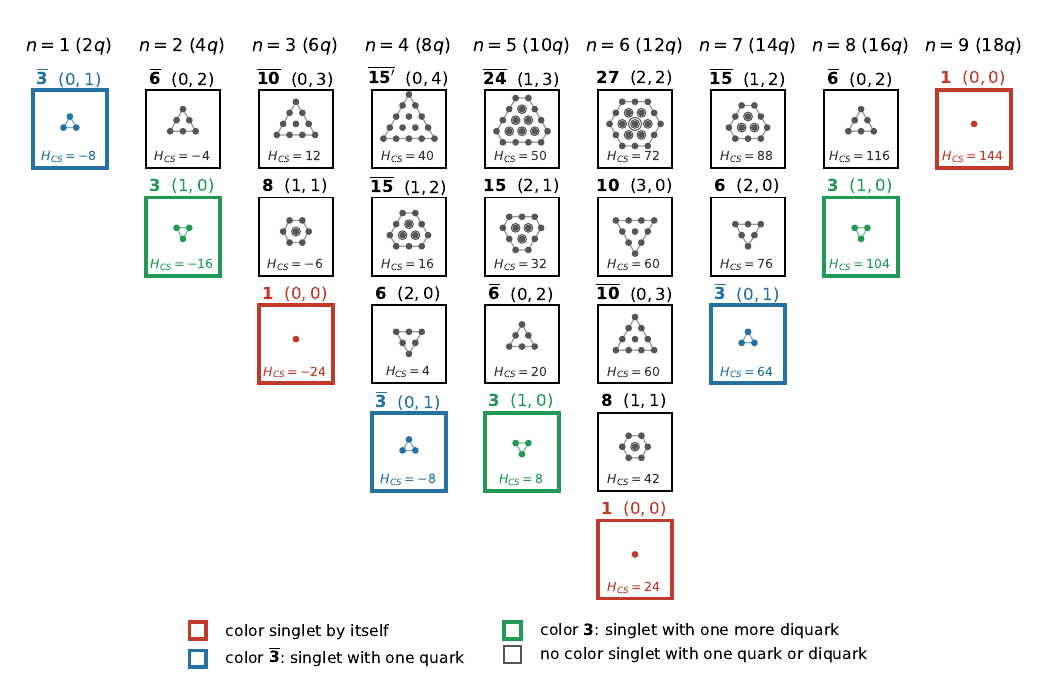}
\caption{Flavor weight diagrams of the $n$-diquark space for flavor SU(3) (Table~\ref{tab:ndiquark}), one column for each $n$. In each diagram the horizontal axis is $I_3$ and the vertical axis is $Y$, with $Y=\frac23$ for the $ud$ diquark and $-\frac13$ for $us$ and $ds$. Rings indicate weight multiplicities. The color representation equals the flavor one, Eq.~\eqref{eq:cauchy}. Red boxes are color singlets by themselves. Blue boxes have color $\threebar$ and form a singlet with one additional quark. Green boxes have color $\bm3$ and form a singlet with one additional diquark. Gray boxes form no color singlet with one additional quark or diquark. Each box gives $H_{CS}$ of Eq.~\eqref{eq:hcs3}, and the colored multiplet of each column is the lowest one, $H^{\rm min}_{CS}$ of Table~\ref{tab:ndiquark}.}
\label{fig:flavor}
\end{figure*}

Figure~\ref{fig:flavor} shows the flavor weight diagrams of these multiplets. Because the color and flavor labels coincide, each flavor multiplet also fixes the color of the medium.
A color singlet built from diquarks alone is therefore necessarily a flavor singlet, and it exists only for $n=3,6,9$. The familiar $\bar{\bm6}$, $\bm8$, $\bm{10}$, and $\bm{27}$ flavor multiplets are colored states of the medium.
With $N=2n$ and $S=0$, Eq.~\eqref{eq:hcs3} becomes $H_{CS}=N(N-10)+6C$ with $C=C_F=C_C$. For each $n$, the lowest multiplet is therefore the one with the smallest Casimir, namely the flavor $\bm1$, $\bm3$, or $\threebar$, according to $n\bmod3$.
These are exactly the multiplets that become color singlets alone ($\bm1$), with one additional quark ($\threebar$), or with one additional diquark ($\bm3$). They give the color-singlet sector of Sec.~\ref{sec:results}.

For flavor SU(2) the diquark is the isoscalar $ud$ diquark, $V=(\threebar_c,\bm 1_f,0)$.
The $n$-diquark space is the single representation $\Sym^n(\threebar)_c$, i.e., $(0,n)$, which exists only for $n\le4$.

\begin{table}[t]
\caption{Content of the $n$-diquark space for flavor SU(3). All states have $S=0$, and the color and flavor Dynkin labels coincide, so a single $(p,q)$ is listed. ``dim'' is the dimension of the space, and $\langle H_{CS}\rangle$ and $H^{\rm min}_{CS}$ are its trace average and lowest value. For $n\le4$, dim $=\binom{n+8}{8}$.}
\label{tab:ndiquark}
\begin{ruledtabular}
\begin{tabular}{clrrr}
$n$ & $(p,q)_C=(p,q)_F$ & dim & $\langle H_{CS}\rangle$ & $H^{\rm min}_{CS}$\\
\hline
1 & (0,1) & 9 & $-8$ & $-8$\\
2 & (0,2), (1,0) & 45 & $-6.4$ & $-16$\\
3 & (0,3), (1,1), (0,0) & 165 & $4.8$ & $-24$\\
4 & (0,4), (1,2), (2,0), (0,1) & 495 & $25.6$ & $-8$\\
5 & (1,3), (2,1), (0,2), (1,0) & 846 & $43.49$ & $8$\\
6 & (2,2), (3,0), (0,3), (1,1), (0,0) & 994 & $67.61$ & $24$\\
7 & (1,2), (2,0), (0,1) & 270 & $85.6$ & $64$\\
8 & (0,2), (1,0) & 45 & $113.6$ & $104$\\
9 & (0,0) & 1 & $144$ & $144$\\
\end{tabular}
\end{ruledtabular}
\end{table}

\subsection{Probes}

The physical space of the medium plus a probe is the span of $D^\dagger\cdots D^\dagger|p\rangle$, where $|p\rangle$ runs over the probe multiplet, namely a quark $(\bm3_c,\bm3_f,\tfrac12)$, a diquark $(\threebar_c,\threebar_f,0)$, three correlated diquarks $(\bm1_c,\bm1_f,0)$, or an octet baryon $(\bm1_c,\bm8_f,\tfrac12)$.
Since all diquarks carry $S=0$, the total spin equals the probe spin.
The space is invariant under SU(3)$_c\times$SU(3)$_f\times$SU(2)$_s$, and its irreducible content is obtained from the dimensions of its weight spaces (Appendix~\ref{app:method}).

\section{Independence of the number of diquarks}
\label{sec:theorem}

Consider first the case in which the map $\Sym^n(V)\otimes P\to\Lambda^{N}(\mathbb C^{18})$, with $P$ the probe multiplet, is injective.
Since the map is equivariant, the image carries the same representations, and $H_{CS}$, being a function of Casimir operators, has the same spectrum on $\Sym^n(V)\otimes P$.
On the latter space, each Casimir operator is quadratic in the generators, which are sums over the constituents (diquarks and probe). Hence
\begin{align}
H_{CS}=\sum_{i=1}^{n}h_{d_i}+h_p+\sum_{i<j}h_{d_id_j}+\sum_{i}h_{d_ip},
\label{eq:cluster}
\end{align}
where $h_{d_id_j}$ and $h_{d_ip}$ are invariant two-cluster operators.
The maximally mixed state on $\Sym^n(V)$ is invariant under U(9). Its one-cluster reduced density matrix is therefore $\bm 1/9$, and its two-cluster reduced density matrix is the normalized projector onto $\Sym^2(V)$, both independent of $n$.
It follows that
\begin{align}
\langle H_{CS}\rangle_{nd}&=-8n+\frac{n(n-1)}{2}\langle h_{dd}\rangle_{\Sym^2V},\\
\langle H_{CS}\rangle_{nd+p}&=\langle H_{CS}\rangle_{nd}+H^p_{CS}+n\,\langle h_{dp}\rangle_{V\otimes P},
\end{align}
so that $\Delta H^{\rm avg}_{CS}$ in Eq.~\eqref{eq:dH} is independent of $n$ and equal to its value for one surrounding diquark.
For the diquark medium alone this gives $\langle H_{CS}\rangle_{nd}=-8n+\frac{24}{5}n(n-1)$, which reproduces Table~\ref{tab:ndiquark} for $n\le4$.

We have checked the injectivity directly by comparing the dimension of the physical space with $\dim[\Sym^n(V)\otimes P]$. For three correlated diquarks ($\dim P=1$) and the octet baryon ($\dim P=16$), the two agree for $n\le3$ (e.g., $144,720,2640$ for the baryon), and they differ from $n=4$ on.
The diquark probe is identical to the medium diquarks, so the total space is $\Sym^{n+1}(V)$ rather than $\Sym^n(V)\otimes V$, and the argument above does not apply to it directly. The medium average gives instead
$\Delta H^{\rm avg}_{CS}=[\langle H_{CS}\rangle_{(n+1)d}-\langle H_{CS}\rangle_{nd}+8]/(2n)=24/5$ for $n\le3$, where $\langle H_{CS}\rangle_{md}$ follows the formula above up to $m=4$.
For the quark probe, the color-singlet, flavor-singlet $S=\frac12$ three-quark state is already Pauli forbidden at $n=1$, and we find
$\dim\mathcal F_n=18\binom{n+8}{8}-2\binom{n+7}{8}$ for $n\le3$ ($160,792,2880$). Comparing the weight multiplicities in color, flavor, and spin, we have verified that the kernel is $\Sym^{n-1}(V)\otimes K$ as a representation, with $K=(\bm1_c,\bm1_f,\frac12)$ the two-dimensional forbidden one-diquark--quark multiplet.
Pauli blocking thus acts only at the two-cluster level for $n\le3$, and the average follows from two traces of Eq.~\eqref{eq:hcs3}.
With $B_n=\langle H_{CS}\rangle_{nd}=-8n+\frac{24}{5}n(n-1)$ and $d_n=\binom{n+8}{8}$, the formal space $\Sym^n(V)\otimes q$ ($18d_n$ states) has the average $B_n+4n$, because the color and flavor cross terms between the quark and the medium are traceless.
On the kernel ($2d_{n-1}$ states) the color and flavor Casimirs are those of $n-1$ diquarks, and only the quark number and spin terms of Eq.~\eqref{eq:hcs3} change, which gives $B_{n-1}+12n-32$. Hence
\begin{align}
\langle H_{CS}\rangle_{nd+q}&=\tfrac{18d_n(B_n+4n)-2d_{n-1}(B_{n-1}+12n-32)}{18d_n-2d_{n-1}}\nonumber\\
&=B_n+\frac{21}{5}n ,
\end{align}
and $\Delta H^{\rm avg}_{CS}=21/5$ exactly for $n=1$--3, although Pauli-forbidden states are present from the first diquark on (Table~\ref{tab:avg}).
The $n$ dependence of the interaction in diquark matter is therefore entirely due to quark-level Pauli blocking beyond the two-cluster level, which sets in at $n=4$.

\section{Results}
\label{sec:results}

\subsection{Average over the physical space}

\begin{table*}[t]
\caption{$\Delta H^{\rm avg}_{CS}$ of Eq.~\eqref{eq:dH} for different probes (rows) in a medium of $n$ diquarks (columns). The numbers in parentheses are $\epsilon_p$ of Eq.~\eqref{eq:eps}, which includes the internal interaction of the probe. A dash means that no state survives the Pauli principle.}
\label{tab:avg}
\begin{ruledtabular}
\begin{tabular}{llcccccccc}
 & probe & 1d & 2d & 3d & 4d & 5d & 6d & 7d & 8d\\
\hline
SU(3)$_F$ & quark & 4.200 & 4.200 & 4.200 & 2.042 & 2.170 & 1.142 & 1.971 & 1.800\\
 & diquark & 4.800 & 4.800 & 4.800 & 3.236 & 3.212 & 2.166 & 2.571 & 2.400\\
 &  & (0.800) & (2.800) & (3.467) & (2.236) & (2.412) & (1.500) & (2.000) & (1.900)\\
 & three corr.\ diquarks & 4.000 & 4.000 & 4.000 & 3.500 & 3.137 & 2.789 & -- & --\\
 &  & (0) & (2.000) & (2.667) & (2.500) & (2.337) & (2.122) & & \\
 & baryon & 4.000 & 4.000 & 4.000 & 2.987 & 2.598 & 2.211 & 2.400 & --\\
 &  & (1.333) & (2.667) & (3.111) & (2.321) & (2.064) & (1.766) & (2.019) & \\
\hline
SU(2)$_F$ & quark & 5.333 & 5.333 & 5.333 & 3.000 & -- & & & \\
 & diquark & 6.000 (2.000) & 6.000 (4.000) & 6.000 (4.667) & -- & -- & & & \\
 & baryon & 5.333 (2.667) & 5.333 (4.000) & 5.333 (4.444) & -- & -- & & & \\
\end{tabular}
\end{ruledtabular}
\end{table*}

\begin{figure*}[t]
\includegraphics[width=0.9\textwidth]{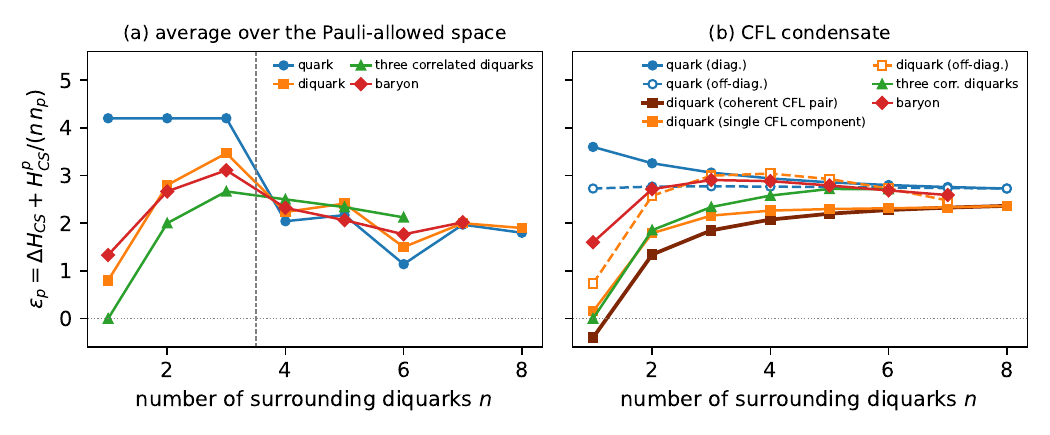}
\caption{Color-spin energy per probe quark and per surrounding diquark including the internal interaction of the probe, $\epsilon_p$ of Eq.~\eqref{eq:eps}, as a function of the number of surrounding diquarks, for flavor SU(3). (a) Trace average over the physical state space. (b) Number-projected CFL condensate. The vertical dashed line marks the end of the trace-average plateau of panel (a). Panel (b) includes the coherent CFL pair, $\sum_ID^\dagger_{\bar I\bar I}$.}
\label{fig:dH}
\end{figure*}

Table~\ref{tab:avg} and Fig.~\ref{fig:dH}(a) show the averaged interaction.
Several features are noteworthy.

(i) For $n\le3$ all entries are constant, as shown in Sec.~\ref{sec:theorem}.
Beyond $n=3$, Pauli blocking reduces $\Delta H^{\rm avg}_{CS}$ for all probes, i.e., the probe--medium interaction becomes less repulsive.
The nonmonotonic behavior at $n=7$ reflects the rapid shrinking of the medium's own state space (Table~\ref{tab:ndiquark}), which enters through the subtraction of $\langle H_{CS}\rangle_{nd}$.

(ii) Without the internal factor, the quark is less repulsive than the baryon once the number of available states starts to fall below the boson count ($n\ge4$, e.g., 2.04 vs 2.99 at $n=4$ and 1.14 vs 2.21 at $n=6$), while the baryon is slightly favored for $n\le3$.
This is the analog, in diquark matter, of the quarkyoniclike behavior found in baryonic matter~\cite{Park:2021hqb,Park:2022jpf}.

(iii) Including the internal interaction, the ordering depends on the filling. For $n\le3$ the three correlated diquarks are the cheapest probe per quark ($\epsilon_p=0$, 2.00, 2.67 for $n=1$--3), followed by the diquark at $n=1$ and by the baryon at $n=2,3$, while the quark is the most expensive.
Once Pauli blocking sets in, the differences between probes shrink to at most one unit, and the quark becomes the cheapest probe at $n=4$ and $6$--$8$ (the baryon at $n=5$).
The internal binding of a composite probe is spread over a growing medium, whereas Pauli blocking reduces the repulsion felt by a single quark.
In an equally weighted medium the short-range color-spin interaction therefore does not favor diquarks at high density. A single quark has the lowest addition cost at most higher fillings, in line with item (ii).

(iv) In the flavor SU(2) case, the medium can hold at most four $ud$
diquarks. A diquark or a baryon cannot be added to four of them, and a
quark cannot be added to five, since five do not exist.

\subsection{Lowest states}

The trace average corresponds to an equal weighting of all states.
As an opposite limit we consider the lowest states,
$\Delta H^{\rm min}_{CS}=[H^{\rm min}_{nd+p}-H^{\rm min}_{nd}-H^p_{CS}]/(n_pn)$, shown in Table~\ref{tab:min}.
Table~\ref{tab:ndiquark} shows that $H^{\rm min}_{nd}=-8n$ for $n\le3$ and $H^{\rm min}_{nd}>-8n$ for $n\ge4$. Hence diquarks gain no further color-spin energy when they are put into a common orbital, and the color-spin interaction alone does not bind diquarks to each other. This does not exclude it as the origin of pairing, since the attraction $H_d=-8$ that forms each good diquark is itself color-spin. What Table~\ref{tab:ndiquark} excludes is an additional attraction between diquarks that would cluster them beyond pairing.
For the lowest states no probe is preferred throughout (Table~\ref{tab:min}). The diquark is the cheapest at $n=2,5,8$, the baryon at $n=3,6$, and the quark at $n=1,4,7$. At $n=1$ the quark is lowest only because the lowest quark--diquark state is colored, namely a color-octet, flavor-singlet three-quark state with $H_{CS}=-14$, below the color-singlet nucleon value $-8$. In the color-singlet sector this exception disappears (Sec.~V.C).

\begin{table}[t]
\caption{$\epsilon_p=\Delta H^{\rm min}_{CS}+H^p_{CS}/(n\,n_p)$ from the lowest states, flavor SU(3).}
\label{tab:min}
\begin{ruledtabular}
\begin{tabular}{lcccccccc}
probe & 1d & 2d & 3d & 4d & 5d & 6d & 7d & 8d\\
\hline
quark & $-6.00$ & 0.00 & 4.00 & 1.50 & 2.40 & 4.00 & 2.57 & 3.00\\
diquark & $-4.00$ & $-2.00$ & 2.67 & 2.00 & 1.60 & 3.33 & 2.86 & 2.50\\
3 corr. & 0 & 2.00 & 2.67 & 3.00 & 3.20 & 3.33 & -- & --\\
baryon & $-2.67$ & 0.67 & 2.44 & 2.33 & 2.67 & 3.22 & 3.05 & --\\
\end{tabular}
\end{ruledtabular}
\end{table}

\subsection{Color-singlet sector}

Neither the average nor the lowest states above are restricted to color-singlet total states.
In our previous works the surroundings were color singlets, so that the color-color interaction between the probe and the medium canceled~\cite{Park:2022qew,Park:2022jpf}.
This is not the case for a colored diquark medium.
To obtain a result free from this issue, we restrict the total system to color singlets, which requires the medium to be in the representation conjugate to the probe color, and define
\begin{align}
\Delta H^{\bm 1}_{CS}=\frac{\langle H_{CS}\rangle_{nd+p}^{C=\bm1}-\langle H_{CS}\rangle_{nd}^{C=\bar p}-H^p_{CS}}{n_p\,n}.
\label{eq:dH1}
\end{align}
This is possible only when the total number of quarks $N=2n+n_p$ is a multiple of three.
Because of the color--flavor correspondence of Eq.~\eqref{eq:cauchy}, the color-singlet sector consists of a single multiplet in every case. It is the flavor octet with $S=\frac12$ for the quark and baryon probes and the flavor singlet with $S=0$ for the diquark and three-correlated-diquark probes.
The average and the lowest state therefore coincide.
The results are given in Table~\ref{tab:singlet}.

\begin{table}[t]
\caption{$\Delta H^{\bm1}_{CS}$ in the total color-singlet sector, Eq.~\eqref{eq:dH1}, for flavor SU(3). $N$ is the total number of quarks, and the last column gives the flavor and spin of the unique color-singlet total multiplet. $\mu_p$ is the energy per added quark, Eq.~\eqref{eq:mu}, with the averages restricted as in Eq.~\eqref{eq:dH1}. The same values are obtained by projecting the CFL condensate onto the color-singlet sector.}
\label{tab:singlet}
\begin{ruledtabular}
\begin{tabular}{cccccc}
$N$ & probe & $n$ & $\Delta H^{\bm1}_{CS}$ & $\mu_p$ & total state\\
\hline
3 & quark & 1 & 0 & 0 & $\bm8$, $S=\frac12$\\
6 & diquark & 2 & 0 & $-4$ & $\bm1$, $S=0$\\
9 & quark & 4 & 3.0 & 12 & $\bm8$, $S=\frac12$\\
9 & baryon & 3 & 4.0 & 9.33 & $\bm8$, $S=\frac12$\\
12 & diquark & 5 & 2.4 & 8 & $\bm1$, $S=0$\\
12 & 3 corr. & 3 & 4.0 & 8 & $\bm1$, $S=0$\\
15 & quark & 7 & 3.43 & 24 & $\bm8$, $S=\frac12$\\
15 & baryon & 6 & 4.0 & 21.33 & $\bm8$, $S=\frac12$\\
18 & diquark & 8 & 3.0 & 20 & $\bm1$, $S=0$\\
18 & 3 corr. & 6 & 4.0 & 20 & $\bm1$, $S=0$\\
\end{tabular}
\end{ruledtabular}
\end{table}

At $N=3$ the quark and one diquark form a nucleon, and $\Delta H^{\bm1}_{CS}=0$ expresses $H_{CS}(N)=H_{CS}(D)=-8$.
Because the total multiplet at a given $N$ is unique, all probes that lead to the same $N$ end in the same state, and comparing them at fixed $N$ compares the starting configurations. These come with different $n$, so we compare the energy per added quark $\mu_p$.
The baryon is cheaper than the quark (9.33 vs 12 at $N=9$, 21.33 vs 24 at $N=15$), and the diquark and the three correlated diquarks cost exactly the same (8 at $N=12$, 20 at $N=18$).
Without the internal factor, the quark is less repulsive than the baryon (3.0 vs 4.0 at $N=9$ and 3.43 vs 4.0 at $N=15$), as in the trace average.

For flavor SU(2), the medium is always in the color representation $(0,n)$, and the only color-singlet case is the nucleon ($n=1$ with a quark probe).

\subsection{2SC and CFL condensates}
\label{sec:condensate}

In a color superconductor the diquarks are not distributed over all states but form a coherent condensate.
In a single orbital, the number-projected condensates are
\begin{align}
|{\rm 2SC},n\rangle&\propto (D^\dagger_{\bar b\bar s})^n|0\rangle,\\
|{\rm CFL},n\rangle&\propto \Big(\sum_{I}D^\dagger_{\bar I\bar I}\Big)^n|0\rangle,
\end{align}
where in the 2SC case the red and green $u,d$ quarks pair, and in the CFL case the color and flavor indices are locked~\cite{Alford:1998mk}.
In the usual notation, $D^\dagger_{\bar a\bar i}$ corresponds to the pairing
$\epsilon^{abc}\epsilon_{ijk}$, so that the 2SC condensate is the
$\epsilon^{\alpha\beta3}\epsilon_{ij3}$ channel and the CFL condensate the sum
over $I$ with equal gaps~\cite{Alford:2007xm}. These are the fixed-number projections of the BCS states restricted to a single orbital. They capture the color--flavor structure of the condensates but not the momentum-space smearing of the Fermi surface.
We compute $\langle H_{CS}\rangle$ directly with Eq.~\eqref{eq:hsecond}. For a probe multiplet, we average over the span of $|{\rm cond},n\rangle\otimes|p\rangle$.
The condensate energies are $\langle H_{CS}\rangle_{\rm 2SC}=-8,-4,12,40$ for $n=1,\dots,4$, and $\langle H_{CS}\rangle_{\rm CFL}=-8,-8.8,-3.43,7.67,24.29,46.29,73.6,106.18,144$ for $n=1,\dots,9$.

\begin{table}[t]
\caption{$\epsilon_p=\Delta H_{CS}+H^p_{CS}/(n\,n_p)$ for probes in the number-projected 2SC condensate. ``$\bar a\bar i$'' denotes the color and flavor indices of a diquark probe $D^\dagger_{\bar a\bar i}$, and $ai$ those of a quark probe.}
\label{tab:2sc}
\begin{ruledtabular}
\begin{tabular}{lcccc}
probe & $n=1$ & 2 & 3 & 4\\
\hline
quark ($ru,rd,gu,gd$) & 6 & 6 & 6 & --\\
quark ($bu,bd$) & 3 & 3 & 3 & 3\\
quark ($s$) & 0 & 0 & 0 & 0\\
diquark ($\bar r\bar u,\bar r\bar d,\bar g\bar u,\bar g\bar d$) & $-1.43$ & 0.50 & 1.07 & 1.25\\
diquark (sharing $\bar b$ or $\bar s$) & 2 & 4 & 4.67 & --\\
three corr.\ diquarks & 0 & 2 & 2.67 & --\\
baryon & 1.58 & 2.91 & 3.33 & 1.00\\
\end{tabular}
\end{ruledtabular}
\end{table}

The 2SC results are shown in Table~\ref{tab:2sc}.
Quarks with the quantum numbers of the paired quarks feel a strong Pauli repulsion and are completely blocked at $n=4$, when the red-green $ud$ orbitals are filled. The unpaired blue $u,d$ quarks and the strange quarks, which remain gapless in the 2SC phase~\cite{Alford:2007xm}, interact only weakly.
Among diquark probes, those carrying $\bar b$ or $\bar s$, i.e., sharing an index with the condensate, are the most expensive. Those whose color and flavor indices both differ from the condensate cost $-1.43$, $0.50$, $1.07$, and $1.25$ for $n=1$--4, less than their two constituent quarks added separately, which cost at least $1.5$ per quark on average. Only the strange quarks, which do not interact with the condensate, are cheaper.
These pairs involve the unpaired blue and strange quarks, and their
pairing, e.g.\ in the $\bar r\bar u$ and $\bar g\bar d$ channels, is compatible with CFL order. This is a statement about addition energies in the prescribed 2SC background, not a calculation of the transition.
We also note that $(D^\dagger_{\bar b\bar s})^n|0\rangle$ is a pure color $(0,n)$ state, so that the 2SC medium cannot form a color singlet with any probe for $n\ge2$. The 2SC results thus describe an intrinsically colored, deconfined medium.

\begin{table*}[t]
\caption{Same as Table~\ref{tab:2sc} for the CFL condensate, $n=1$--$8$. ``Diagonal'' quarks are $ru$, $gd$, $bs$, and a single CFL component is one of $D^\dagger_{\bar r\bar u},D^\dagger_{\bar g\bar d},D^\dagger_{\bar b\bar s}$, averaged over the three. The coherent CFL pair $\sum_ID^\dagger_{\bar I\bar I}$ adds one pair to the condensate and costs $(E_{n+1}-E_n)/(2n)$ with $E_n=\langle H_{CS}\rangle_{\rm CFL}$. These two rows are given to three decimals to resolve their difference at $n=7$.}
\label{tab:cfl}
\begin{ruledtabular}
\setlength{\tabcolsep}{2pt}
\begin{tabular}{lcccccccc}
probe & 1 & 2 & 3 & 4 & 5 & 6 & 7 & 8\\
\hline
quark (diag.) & 3.60 & 3.26 & 3.06 & 2.94 & 2.86 & 2.80 & 2.76 & 2.73\\
quark (off-diag.) & 2.73 & 2.77 & 2.77 & 2.77 & 2.76 & 2.75 & 2.74 & 2.73\\
diquark (coherent CFL pair) & $-0.400$ & 1.343 & 1.850 & 2.077 & 2.200 & 2.276 & 2.327 & 2.364\\
diquark (single CFL component) & 0.154 & 1.791 & 2.160 & 2.265 & 2.297 & 2.313 & 2.333 & 2.364\\
diquark (off-diag.) & 0.74 & 2.58 & 3.00 & 3.04 & 2.93 & 2.73 & 2.48 & --\\
3 corr. & 0 & 1.85 & 2.34 & 2.58 & 2.72 & 2.71 & -- & --\\
baryon & 1.60 & 2.72 & 2.91 & 2.88 & 2.79 & 2.69 & 2.59 & --\\
\end{tabular}
\end{ruledtabular}
\end{table*}

The CFL results are shown in Table~\ref{tab:cfl} and Fig.~\ref{fig:dH}(b).
In contrast to the trace average, the interaction varies smoothly with $n$. Among the other probes the diquark in a single CFL component is the cheapest for $n\ge2$, by 0.06--0.42 below the next one, and its value stays near 2.3.
The coherent CFL pair $\sum_ID^\dagger_{\bar I\bar I}$, which grows the condensate itself, is cheaper still for $n\le7$ and coincides with the single CFL component at $n=8$, $26/11$. It is the cheapest probe at every $n$, and at $n=1$ it costs $-0.40$, below the three correlated diquarks.
This does not contradict Table~\ref{tab:ndiquark}. Two CFL pairs have $E_2=-8.8$, below one pair ($-8$) but above two isolated pairs ($-16$), so a low addition energy to the condensate is not a binding between diquarks.
All quark probes reach the same value, $30/11\simeq2.73$, at $n=8$. There seventeen quarks in the eighteen $s$-shell states form a single hole, whose quantum numbers fix $H_{CS}=128$ for every quark type. The residual SU(3)$_{c+f}$, under which the quarks form $\bm8\oplus\bm1$, only makes the octet quarks degenerate. The diagonal quarks contain the singlet and differ from the off-diagonal ones for $n<8$.
A prescribed coherent CFL medium thus favors adding pairs of its own type, in contrast to the equally weighted medium of Sec.~\ref{sec:results}.
When the CFL condensate is projected onto the sector that forms a total color singlet with the probe, only the single-CFL-component diquarks survive among the diquark probes, and the values of Table~\ref{tab:singlet} are reproduced exactly, because the color-singlet sector is unique.
The CFL condensate thus connects smoothly to the color-singlet result.

\section{Sensitivity to the model assumptions}
\label{sec:robust}

\subsection{Axial-vector diquark}

So far the medium and the diquark probe were scalar good diquarks. As a check, we take the axial-vector diquark $(\threebar_c,\bm6_f,S=1)$, with $H^p_{CS}=8/3$, as a probe.
We find $\Delta H^{\rm avg}_{CS}=4,4,3.14,1.91,1.51,1.22,1.74$ and $\epsilon_p=5.33,4.67,3.58,2.24,1.78,1.44,1.93$ for $n=1,\dots,7$.
For $n\le3$ this is far above the good diquark and even above the quark.
Because its multiplet is six times larger, the axial-vector diquark is Pauli blocked earlier, already at $n=3$ ($\dim\mathcal F_3=7020<8910$).
For $n\ge4$ its repulsive internal factor is diluted, and it becomes comparable to the good diquark, with $\epsilon_p=2.24$ for both at $n=4$ and slightly lower values for $n\ge5$.
As in free space and in baryonic matter~\cite{Jaffe:2004ph,Park:2022qew}, the good diquark is thus the favored diquark in a dilute diquark medium, whereas near saturation the color-spin interaction no longer distinguishes the two.

\subsection{Color-electric and three-quark interactions}

In a colored medium, two further parts of a constituent-quark Hamiltonian contribute to the probe--medium interaction, the color-electric two-body term and the three-quark force.
In a common orbital each term is a spatial coefficient times a color factor, so the factors must be weighted by their coefficients before they are compared with $H_{CS}$.
We use a quark-model Hamiltonian with color-Coulomb, linear, and constant color-electric terms, a smeared color-spin term, and the three-quark force of Ref.~\cite{Noh:2024nvp,Baek:2026uup}, with all parameters fitted to meson and baryon masses.
For light quarks in a harmonic-oscillator orbital of size $b=0.5$ fm, its color-spin term equals $\kappa H_{CS}$ with $\kappa=18.0$ MeV, the value used in Sec.~\ref{sec:orbital}.

The color-electric term then equals $c_{E}\sum_{i<j}\lambda^c_i\cdot\lambda^c_j$, with $\sum_{i<j}\lambda^c_i\cdot\lambda^c_j=2C_C-8N/3$ and $c_E\simeq62$ MeV. The constant term of the potential dominates $c_E$, which varies from 82 to 25 MeV for $b=0.45$--$0.6$ fm.
For equal masses, the color-color part of the three-quark force becomes
$c_3(N-2)\sum_{i<j}\lambda^c_i\cdot\lambda^c_j+c_d\sum_{i<j<k}d^{abc}F^a_iF^b_jF^c_k$, with $c_3\simeq-2.3$ MeV, $c_d\simeq-81$ MeV, and
$\sum_{i<j<k}d^{abc}F^a_iF^b_jF^c_k=[C^{(3)}_C-10N/9-\frac52(C_C-4N/3)]/6$, where $C^{(3)}_C=(p-q)(2p+q+3)(p+2q+3)/18$ is the cubic Casimir.
Its spin-dependent parts are an order of magnitude smaller in this model and are omitted.
The color-electric and three-quark contributions do not cancel between a colored probe and a colored medium. We define their probe--medium parts as in Eq.~\eqref{eq:dH}.

In the trace average for $n\le3$, they add about $10$ MeV per probe quark for the quark, $36$ MeV for the diquark, and $18$ MeV for the baryon and the three correlated diquarks, compared with $72$--$86$ MeV from the color-spin term.
For $n\ge4$ the color-electric part becomes attractive, between $-10$ and $-60$ MeV, and is comparable to the color-spin part.
Including all terms and the internal color-spin energy as in Eq.~\eqref{eq:eps}, the three correlated diquarks remain the cheapest probe for $n\le3$, with $18$, $54$, and $65$ MeV per quark for $n=1$--3, followed by the baryon. The color-electric repulsion between the colored diquark and the colored medium raises the diquark to the level of the quark.
For $n\ge4$ the color-electric attraction of a colored probe makes the quark the cheapest probe, with $-17$ to $15$ MeV per quark, ahead of the diquark with $9$ to $31$ MeV.
These numbers are estimates. The trace average contains colored total states, for which the color-electric interaction, in particular its constant part, is not well defined, and the coefficients depend on $b$.
They use the coefficients at $b=0.5$ fm for all $n$ with the normalization of Eq.~\eqref{eq:eps}. Along a fixed-density path the color-electric coefficients change differently from $\kappa$, so that the color-electric parts for large $n$ indicate signs and orders of magnitude only.

In the color-singlet sector, the color-electric term binds a colored probe to the medium, e.g.\ by about $-330$ MeV per quark when a quark and one diquark form a nucleon. This is the confining attraction that neutralizes the probe color, not a short-range probe--medium interaction.
For the color-singlet probes the color-electric part vanishes, and the three-quark force adds the same $24$ MeV per quark to the baryon and to the three correlated diquarks, so their order is unchanged.
The color-electric and three-quark terms thus reinforce the ordering obtained from $H_{CS}$, with the three correlated diquarks cheapest for $n\le3$ and the quark for $n\ge4$. For colored probes at $n\le3$, however, the color-electric interaction is of the same order as the differences between probes.

\subsection{Flavor SU(3) breaking}

With $m_s\neq m_u$, Eq.~\eqref{eq:vcs} is no longer a function of Casimir operators. We evaluate it as $H_{CS}=-\frac12\sum_{ak}\tilde Q^{ak}\tilde Q^{ak}+8\sum_f w_f^2N_f$ with $\tilde Q^{ak}=\sum_f w_f\,c^\dagger_f\lambda^a\sigma^kc_f$ and $w_f=m_u/m_f$. The state spaces are those of the symmetric limit, built from the scalar
diquark operators of Eq.~\eqref{eq:diquark}.
For $m_u/m_s=0.6$, we obtain the values in Table~\ref{tab:mass}.
All entries are reduced by about $25\%$, and the $n$ independence for $n\le3$ survives, as expected from the cluster argument of Sec.~\ref{sec:theorem}, which requires only a two-body interaction and an invariant medium.
The ordering is that of the symmetric limit. For $n\le3$ the three correlated diquarks are the cheapest probe and the quark the most expensive, and for $n\ge4$ the quark ($n=4$) or the baryon ($n=5$) is the cheapest.

\begin{table}[t]
\caption{Same as Table~\ref{tab:avg} (SU(3)$_F$) with $m_u/m_s=0.6$. The values of $\epsilon_p$ are in parentheses. The probe internal factors are averaged over the multiplet, giving $\langle H^p\rangle=-5.87$ (diquark), $-5.76$ (baryon), and $-17.12$ (three correlated diquarks).}
\label{tab:mass}
\begin{ruledtabular}
\begin{tabular}{lccccc}
probe & 1d & 2d & 3d & 4d & 5d\\
\hline
quark & 3.197 & 3.197 & 3.197 & 1.595 & 1.690\\
diquark & 3.643 & 3.643 & 3.643 & 2.482 & 2.464\\
 & (0.709) & (2.176) & (2.665) & (1.749) & (1.877)\\
3 corr. & 3.049 & 3.049 & 3.049 & 2.678 & 2.408\\
 & (0.196) & (1.622) & (2.098) & (1.965) & (1.837)\\
baryon & 3.049 & 3.049 & 3.049 & 2.297 & 2.006\\
 & (1.129) & (2.089) & (2.409) & (1.817) & (1.622)\\
\end{tabular}
\end{ruledtabular}
\end{table}

\subsection{Orbital excitation and the energy scale}
\label{sec:orbital}

The assumption of a single common orbital is the most severe limitation of the present work. To test it, we have diagonalized $H_{CS}$ with a contact interaction in the space of the harmonic-oscillator $0s$ and $0p$ orbitals, with the center-of-mass motion removed. We omit the coupling between configurations whose numbers of $p$ quarks differ by two, such as $0s^N$ and $0s^{N-2}0p^2$, which have the same parity and can mix. The details are given in Appendix~\ref{app:pwave}.

To set the energy scale, we note that $H_{CS}(\Delta)-H_{CS}(N)=16$, so that for baryon-size wave functions one unit of $H_{CS}$ corresponds to $\kappa\simeq(M_\Delta-M_N)/16\simeq18$ MeV.
A quark and one surrounding diquark form a three-quark system of baryon size, so we identify this value with $\kappa_*=\kappa(n=1)$ of Sec.~\ref{sec:model}.
On the fixed-density path the energies per quark are $\kappa_*\epsilon_p$, and the differences between probes in Tables~\ref{tab:avg}--\ref{tab:cfl} are at most a few tens of MeV per quark, e.g., $3.4\kappa_*\simeq60$ MeV between a quark and a diquark at $n=1$ and at most one unit, $\simeq18$ MeV, between all probes for $n\ge4$. At a fixed orbital size the same differences are multiplied by $n$.
Promoting a quark to $0p$ costs the oscillator gap $\hbar\omega=(\hbar c)^2/(m_ub^2)\simeq0.3$--$0.5$ GeV for an orbital size $b\simeq0.5$--$0.6$ fm.
The orbital-excitation calculation below and in Appendix~\ref{app:pwave} keeps $b$, $\kappa=18$ MeV, and $\hbar\omega$ fixed for all $n$. It describes a local region of fixed size, in which the density grows with $n$, and not the fixed-density path. Along that path $b$ grows as $n^{1/3}$, so that $\kappa\propto b^{-3}$ and $\hbar\omega\propto b^{-2}$ both decrease, e.g., to $2.25$ MeV and $0.125$ GeV at $n=8$ when $b=0.5$ fm and $\hbar\omega=0.5$ GeV at $n=1$. The ratio $\kappa/\hbar\omega\propto b^{-1}$ then decreases as $n^{-1/3}$, so that orbital excitation is less favorable at fixed density than in the fixed-size calculation.

A single probe quark in $0p$ interacts only with the spatially symmetric part of each pair and is not Pauli blocked. Its interaction with the medium is $\Delta H^{\rm avg}_{CS}=8/9$ for all $n$, or $0.18$--$0.80$ for $n=1$--$4$ after the center-of-mass excitation is removed, much smaller than the $0s$ value $4.2$.
For $n\le3$ and for the color-singlet states up to $N=12$, occupying $0p$ lowers the color-spin energy by at most about 12 units, well below $\hbar\omega/\kappa$.
The occupation of higher orbitals is therefore governed by the capacity of the $s$ shell rather than by the color-spin interaction, and the results for $n\le3$ are unaffected.
Near $s$-shell saturation, $0p$ occupation can become favorable for $\hbar\omega\lesssim0.40$--$0.45$ GeV, so that the entries for $n=6$--$8$ depend on $\hbar\omega$.
The same calculation shows that the $n$ independence of Sec.~\ref{sec:theorem} is exact only for probes that occupy the medium orbital fully or not at all, and that the color--flavor correspondence of Eq.~\eqref{eq:cauchy} and the uniqueness of the color-singlet sector are properties of the $0s$ space.
Once quarks occupy higher orbitals, they are nearly decoupled from the diquarks in the lowest orbital. This is reminiscent of the quarkyonic momentum-space structure, with the roles of baryons played by diquarks.
A full treatment of several shells, with $\kappa$ and $\hbar\omega$ tied to the density, is left for future work.

\section{Bosonic interpretation}
\label{sec:boson}

The structure found above has a compact bosonic interpretation.
Treat each good diquark as an elementary boson, i.e., as the fundamental representation of U(9) with $\bm 9=\threebar_c\otimes\threebar_f$. The $n$-diquark states then form the totally symmetric representation $[n]$ of U(9), and the branching U(9)$\supset$U(3)$_c\otimes$U(3)$_f$ is the Cauchy identity~\eqref{eq:cauchy}.
This construction is analogous to the interacting boson model of nuclei, with the quark-level Pauli principle playing the role of the Otsuka--Arima--Iachello boson mapping~\cite{Otsuka:1978zz}.
The diquark--diquark interaction obtained from the four-quark problem is exactly two-body, $h_{ij}=8+6\,(2\bm T_i\cdot\bm T_j)$, where $\bm T_i$ is the color generator of boson $i$, the sum of $\lambda^c/2$ over its two quarks, so that $\bm T_i^2=C(\threebar)=\frac43$. Summing it gives
\begin{align}
H_B(\lambda)=4n^2-20n+6\,C_C(\lambda),
\end{align}
which coincides with Eq.~\eqref{eq:hcs3} for $N=2n$, $S=0$, $C_F=C_C$.
The boson and quark descriptions therefore differ only in which states survive the antisymmetrization.
We find that the component $\lambda$ survives if and only if
\begin{align}
\lambda\subseteq[4,2,2]\quad\text{or}\quad\lambda\subseteq[3,3,3]
\qquad(\lambda_1\le4,\ \lambda_1+\lambda_2\le6),
\label{eq:paulirule}
\end{align}
in agreement with Table~\ref{tab:ndiquark}.
In boson language, Eq.~\eqref{eq:paulirule} is a five-body plus seven-body hard-core constraint generated by the components $[5]$ and $[4,3]$.
The component $[5]$ describes five diquarks in the same color--flavor direction. These cannot be accommodated because a single 2SC-type block (two colors $\times$ two flavors $\times$ two spins) holds at most four diquarks.
This saturation is the origin of the onset of Pauli blocking at $n=4$ in Table~\ref{tab:avg}. Without the constraint, the boson model would give an $n$-independent $\Delta H^{\rm avg}_{CS}$ for all $n$.
For $\Omega$ degenerate orbitals, with diquarks
$D^\dagger=\sum_{o=1}^{\Omega}D^\dagger_o$ that are symmetric in the orbital
index, the bounds in Eq.~\eqref{eq:paulirule} are multiplied by $\Omega$,
and the onset moves to $n=4\Omega$~\cite{Park:paperII}. The onset is therefore a local condition of four diquarks per orbital, not an artifact of the single-orbital model.
The bosonic formulation, its norm kernel and condensates, the extension to axial-vector diquarks, and the dynamical effects of quark hopping between neighboring regions, including the emergence of color-octet baryon pairs, will be discussed in forthcoming work~\cite{Park:paperII}.

\section{Summary}
\label{sec:summary}

We have studied the color-spin interaction of a quark, a diquark, three correlated diquarks, and an octet baryon in a medium of scalar diquarks in a constituent quark model, constructing the physical state space explicitly from diquark operators.
Our main findings are as follows.

\begin{itemize}
\item For bosonic diquarks the color and flavor representations of the medium coincide [Eq.~\eqref{eq:cauchy}]. Selecting states by separate outer products in color and flavor overcounts the state space.
\item The averaged interaction per surrounding diquark is independent of $n$ as long as the diquarks behave as elementary bosons. Its departure from this plateau comes entirely from the quark substructure of the diquarks, which limits how many of them fit into one orbital and becomes effective at $n=4$. For the quark probe the plateau holds exactly although Pauli-forbidden states are present from $n=1$, because the kernel is $\Sym^{n-1}(V)\otimes K$ for $n\le3$.
\item Without the internal factor, the quark is less repulsive than the baryon at high density. Including the internal interaction of the probe with the same density scaling, the three correlated diquarks are the cheapest probe per quark for $n\le3$ and the quark is the most expensive. After the onset of Pauli blocking the differences shrink, and the quark becomes the cheapest probe in the trace average for most $n$. The lowest states show no uniform ordering. In the color-singlet sector the diquark and the three correlated diquarks cost the same per added quark, and the baryon less than the quark. In the number-projected CFL condensate the coherent CFL pair is the cheapest probe for every $n$, and among the other probes the diquark in a single CFL component for $n\ge2$.
\item The color-spin interaction alone does not bind diquarks to each other.
\item In the 2SC medium, unpaired quarks interact weakly, and the diquarks that pair the unpaired blue and
strange quarks cost less than these quarks added separately. The addition energies thus favor pairing channels compatible with CFL order in the prescribed 2SC background.
\item These patterns are stable against the main simplifications (Sec.~\ref{sec:robust}). The axial-vector diquark is much more expensive than the good diquark for $n\le3$ and comparable to it near saturation. Weighted by their coefficients in a fitted quark model, the color-electric and three-quark terms add tens of MeV per probe quark and reinforce the ordering, with the three correlated diquarks cheapest for $n\le3$ and the quark for $n\ge4$. With $m_u/m_s=0.6$, all values shrink by about $25\%$ and the ordering is unchanged. A quark in the $0p$ orbital interacts only weakly with the diquark medium, and in the $0s+0p$ sectors with a fixed number of $p$ quarks examined here, orbital excitation becomes competitive only near $s$-shell saturation for $\hbar\omega\lesssim0.45$ GeV.
\end{itemize}

Whether diquarks are favored in dense matter is therefore not decided by the short-range color-spin interaction alone. In an equally weighted diquark medium, six-quark correlations are favored at low filling, and after the onset of Pauli blocking a single quark has the lowest addition cost at most fillings, a quarkyoniclike pattern. A prescribed coherent CFL-type condensate, by contrast, favors adding pairs of its own type at every filling, which is compatible with CFL order, although the present calculation does not determine the condensate self-consistently. The color-spin interaction forms the pairs but does not bind them to each other, so whether the pairs condense is decided by the coherence of the medium and by the other parts of the interaction, not by a diquark--diquark attraction.
The most important extension left for future work is a multi-shell treatment in which the spatial scale is tied to the density, including mixed media of scalar and axial-vector diquarks and the strength of three-body forces~\cite{Noh:2024nvp,Baek:2026uup}.
It will also be interesting to combine the present interaction factors with dynamical simulations of multiquark clustering in neutron-star matter~\cite{Yasutake:2026fjh}.

\begin{acknowledgments}
This work was supported by the Korea National Research Foundation under the grant No. RS-2025-23963552.
\end{acknowledgments}

\appendix

\section{Numerical construction of the state space}
\label{app:method}

We represent states in the Fock space of the 18 single-particle orbitals by occupation bit strings.
The diquark operator of Eq.~\eqref{eq:diquark} and the probe creation operators are applied to the vacuum.
The octet-baryon multiplet is obtained by diagonalizing $\bm S^2$ in the space of color-singlet three-quark states, $\epsilon_{abc}c^\dagger_{ax}c^\dagger_{by}c^\dagger_{cz}|0\rangle$, which forms the $\bm{56}$ of SU(6), and keeping $S=\frac12$.
The three-correlated-diquark state is $\det(D^\dagger)|0\rangle$, the totally antisymmetric combination $\sum_\pi{\rm sgn}(\pi)D^\dagger_{\bar 1\overline{\pi(1)}}D^\dagger_{\bar 2\overline{\pi(2)}}D^\dagger_{\bar 3\overline{\pi(3)}}|0\rangle$.

Because the physical space $\mathcal F$ is invariant under $G={\rm U}(3)_c\times{\rm U}(3)_f\times{\rm U}(2)_s$, its content is determined by the dimensions $m(\mu)$ of its weight spaces. Here the weight $\mu$ lists the numbers of quarks of each color, flavor,
and spin, and $m(\mu)$ is the number of linearly independent states with
this weight.
We compute $m(\mu)$ as the rank of the set of vectors
$D^\dagger_{\bar a_1\bar i_1}\cdots D^\dagger_{\bar a_n\bar i_n}|p\rangle$ with total weight $\mu$,
where the multisets $\{(\bar a_k,\bar i_k)\}$ are enumerated as $3\times3$
contingency tables with fixed margins.
The multiplicity of the irreducible representation with highest weight $\lambda$ is then
\begin{align}
n_\lambda=\sum_{w\in W}{\rm sgn}(w)\,m\big(\lambda+\delta-w\delta\big),
\end{align}
where $W=S_3\times S_3\times S_2$ is the Weyl group and $\delta$ is half the sum of positive roots, $\delta=(2,1,0)$ for each U(3) and $(1,0)$ for U(2).
Since $H_{CS}$ is a function of Casimir operators [Eq.~\eqref{eq:hcs3}], the trace average follows directly from $\{n_\lambda\}$.
We checked that the method reproduces the known values $H_{CS}=-8$ (diquark, octet baryon), $-24$ (three correlated diquarks), and the full dimensions $\binom{n+8}{8}$ for $n\le4$.

For the condensates of Sec.~\ref{sec:condensate}, which are not eigenstates of the Casimir operators, we evaluate Eq.~\eqref{eq:hsecond} with sparse matrices.
Color projections use the polynomial projector
$\prod_{\rho\neq\lambda}(C_C-c_\rho)/(c_\lambda-c_\rho)$ over the color Casimir values $c_\rho$ allowed for $N$ quarks. The triality of $N$ distinguishes $(1,0)$ from $(0,1)$.
As a check, $\langle H_{CS}\rangle$ for the trace average over all two-diquark states obtained with Eq.~\eqref{eq:hsecond} agrees with Table~\ref{tab:ndiquark}.

\section{Calculation in the $0s+0p$ space}
\label{app:pwave}

This appendix describes the calculation quoted in Sec.~\ref{sec:orbital}.

\paragraph*{A single probe quark in $0p$.}
We first place the $2n$ quarks of the medium in $0s$ and the probe quark in $0p$.
For a contact interaction, $\langle sp|\delta|sp\rangle=\langle sp|\delta|ps\rangle=\frac12\langle ss|\delta|ss\rangle$, so that an $s$--$p$ pair acts with $\frac12O_{ij}(1-P^{CFS}_{ij})$, where $O_{ij}=-\lambda^c_i\cdot\lambda^c_j\,\bm\sigma_i\cdot\bm\sigma_j$. Only the spatially symmetric part of the pair interacts.
The probe in $0p$ is not Pauli blocked by the medium. Using ${\rm Tr}_p\,O_{ip}=0$ and ${\rm Tr}_p\,(O_{ip}P^{CFS}_{ip})=-16$, the trace average gives exactly
\begin{align}
\Delta H^{\rm avg}_{CS}(0p\ {\rm quark})=\frac{8}{9}
\end{align}
for all $n$, which we have confirmed numerically for $n\le4$.
The state with a single $0p$ quark contains a spurious center-of-mass excitation. Projecting the space onto the subspace orthogonal to $R^\dagger|(0s)^N\rangle$, with $R^\dagger=\sum_\alpha c^\dagger_{p\alpha}c_{s\alpha}$, gives $\Delta H^{\rm avg}_{CS}=0.18,0.55,0.71,0.80$ for $n=1$--$4$. The value approaches $8/9$ as the medium grows, since the spurious
center-of-mass component of a single-quark excitation decreases as $1/N$.
At a fixed orbital size with $\kappa=18$ MeV, as in the rest of this appendix, promoting the probe quark from $0s$ to $0p$ thus gains $n(4.2-0.89)\kappa\simeq60n$ MeV of color-spin energy for $n\le3$, and only $\simeq20n$ MeV for $n\ge4$, both below $\hbar\omega$ for $n$ below the filling of the $s$ shell.

\paragraph*{Full $0s+0p$ space.}
The single-particle space consists of the harmonic-oscillator $0s$ and $0p$ ($m=0,\pm1$) orbitals, each with three colors, three flavors, and two spins, which gives 72 states.
The interaction is the contact color-spin interaction $\sum_{i<j}\delta(\bm r_{ij})\,O_{ij}$ with the spatial integrals of oscillator wave functions, normalized so that it reduces to $H_{CS}$ in the $(0s)^N$ space.
States are classified by the number $k$ of quarks in $0p$.
We keep $k\le2$, and $k=3$ (and $k=4$ for $N=16$--$18$) for the lowest states of each spatial symmetry, and we drop the coupling between configurations that differ by two quanta, as in the single-orbital treatment of the main text.
Spurious center-of-mass excitations are removed by keeping only the null space of the center-of-mass lowering operator $B_m\propto\sum_\alpha c^\dagger_{s\alpha}c_{p_m\alpha}$. For $k=1$ this coincides with the projection used above.
For $k\ge2$ our model space omits the $1s$ and $0d$ orbitals, i.e., states in which one quark carries two quanta.
These intrinsic $2\hbar\omega$ states, with spatial diagrams $[N]$ or $[N-1,1]$, have the color--flavor--spin symmetry of states with fewer quanta.
For a contact interaction their color-spin energy is somewhat lower, by up to 8.5 units for $N\le18$, because radial excitation of the relative motion increases the contact density.
A calculation in the complete $2\hbar\omega$ space shows that they lower the minima by at most 0.2 units for $N\ge6$ and leave the threshold values of $\hbar\omega$ discussed below unchanged within 2 MeV.
The $(0s)^N$ spectra reproduce Eq.~\eqref{eq:hcs3} in every sector, and the projected $0p$-quark values $0.18,0.55,0.71$ given above are reproduced.

Two quarks in a relative $P$ wave have a symmetric color--flavor--spin wave function, $(\threebar_c,\bm6_f,S=0)$ or $(\threebar_c,\threebar_f,S=1)$ for color $\threebar$.
Their contact color-spin energy vanishes, because the relative wave function vanishes at contact. A $P$-wave diquark therefore costs $\hbar\omega$ and loses the full $-8$ of the good diquark.
Two $P$-wave diquarks coupled to $L=0$ give at best $H_{CS}=-5.85$ for $N=4$, compared with $-16$ for two good diquarks.

Table~\ref{tab:pwave} lists the lowest intrinsic $H_{CS}$ for $k=0,1,2$ and the excitation energy $\Delta E=\kappa(H_k-H_0)+k\hbar\omega$ of the lowest $p$-occupied state.
For $k=0$ and even $N$, the minima are those of Table~\ref{tab:ndiquark}.
For $n\le3$ and for the color-singlet states up to $N=12$, the color-spin gain from $0p$ occupation is at most about 12 units, and the six-quark color singlet gains only 0.15 units.
For $N\le12$, $\Delta E$ is positive for $\hbar\omega\ge0.3$ GeV. Including all spatial Young diagrams with up to three quanta, the values of $\hbar\omega$ below which $p$ occupation pays are at most 0.27 GeV for $N\le12$, and lie at 0.29--0.41 GeV for $N\ge13$. Four quanta give 0.38--0.39 GeV for $N=16$--$18$, so the gain per quantum does not grow with $k$.
With $\hbar\omega=0.3$ GeV, $p$ occupation thus becomes favorable from $N=13$, where $\Delta E$ first turns negative, while with $\hbar\omega=0.5$ GeV it never does up to $N=18$.
The lowest $p$-occupied states near saturation belong to the two-row spatial Young diagrams $[N-k,k]$, i.e., a spatially symmetric group of $N-k$ quarks in $0s$ and one of $k$ quarks in $0p$, rather than to $P$-wave pairs.
Replacing the contact interaction by a Gaussian with a range equal to the oscillator length gives the $(\threebar_c,\bm6_f,S=0)$ $P$-wave diquark $H_{CS}=-2.7$, leaves the paired-$P$ configuration far above two good diquarks, and raises the threshold values of $\hbar\omega$ by 10--40 MeV. For $N=11$ the threshold rises to 0.30 GeV, so that with $\hbar\omega=0.3$ GeV the onset moves to $N\approx11$.
Compression of the medium shrinks the orbital. The contact strength scales as $\kappa\propto b^{-3}$ and the gap as $\hbar\omega\propto b^{-2}$, so compression favors $p$ occupation near saturation.
Conversely, along the fixed-density path of Sec.~\ref{sec:model}, where $b\propto n^{1/3}$, the ratio $\kappa/\hbar\omega$ decreases as $n^{-1/3}$, and the fixed-size thresholds above overestimate the tendency to $p$ occupation.

The same calculation qualifies two statements of the main text. The $n$ independence of Sec.~\ref{sec:theorem} is exact for probes that occupy the medium orbital fully or not at all, whereas a probe with quarks in both $0s$ and $0p$, such as a $P$-wave diquark, varies by 3--7\% between $n=1$ and 3. States that break the color--flavor correspondence or the uniqueness of the color-singlet sector lie at least $\hbar\omega$ above the $0s$ ground state for $n\le6$.

\begin{table}[t]
\caption{Lowest intrinsic $H_{CS}$ (center-of-mass excitation removed) of $N$ quarks
with $k$ quarks in $0p$, and the excitation energy $\Delta E=\min_{k=1,2}[\kappa(H_k-H_0)+k\hbar\omega]$ (MeV) of the lowest state with at least one quark in $0p$, for $\hbar\omega=0.3$ and $0.5$ GeV,
with $\kappa=18$ MeV. Negative $\Delta E$ means that $p$ occupation is favored.}
\label{tab:pwave}
\begin{ruledtabular}
\begin{tabular}{rrrrrr}
$N$ & $k=0$ & $k=1$ & $k=2$ & $\Delta E_{0.3}$ & $\Delta E_{0.5}$\\
\hline
4 & $-16$ & $-19.53$ & $-18.73$ & 236 & 436\\
6 & $-24$ & $-24.09$ & $-31.75$ & 298 & 498\\
7 & $-12$ & $-24.43$ & $-29.12$ & 76 & 276\\
8 & $-8$ & $-15.79$ & $-33.63$ & 139 & 360\\
10 & 8 & $-4.48$ & $-18.89$ & 75 & 275\\
12 & 24 & 17.71 & $-1.24$ & 146 & 387\\
13 & 48 & 27.60 & 14.15 & $-67$ & 133\\
14 & 64 & 49.22 & 21.76 & $-160$ & 234\\
16 & 104 & 85.12 & 61.34 & $-168$ & 160\\
17 & 128 & 108.45 & 81.99 & $-228$ & 148\\
18 & 144 & 130.33 & 103.36 & $-132$ & 254\\
\end{tabular}
\end{ruledtabular}
\end{table}

\bibliography{refs}

@article{Demorest:2010bx,
    author = "Demorest, Paul and Pennucci, Tim and Ransom, Scott and Roberts, Mallory and Hessels, Jason",
    title = "{Shapiro Delay Measurement of A Two Solar Mass Neutron Star}",
    eprint = "1010.5788",
    archivePrefix = "arXiv",
    primaryClass = "astro-ph.HE",
    doi = "10.1038/nature09466",
    journal = "Nature",
    volume = "467",
    pages = "1081--1083",
    year = "2010"
}

@article{Antoniadis:2013pzd,
    author = "Antoniadis, John and others",
    title = "{A Massive Pulsar in a Compact Relativistic Binary}",
    eprint = "1304.6875",
    archivePrefix = "arXiv",
    primaryClass = "astro-ph.HE",
    doi = "10.1126/science.1233232",
    journal = "Science",
    volume = "340",
    pages = "6131",
    year = "2013"
}

@article{NANOGrav:2019jur,
    author = "Cromartie, H. T. and others",
    collaboration = "NANOGrav",
    title = "{Relativistic Shapiro delay measurements of an extremely massive millisecond pulsar}",
    eprint = "1904.06759",
    archivePrefix = "arXiv",
    primaryClass = "astro-ph.HE",
    doi = "10.1038/s41550-019-0880-2",
    journal = "Nature Astron.",
    volume = "4",
    number = "1",
    pages = "72--76",
    year = "2019"
}

@article{LIGOScientific:2017vwq,
    author = "Abbott, B. P. and others",
    collaboration = "LIGO Scientific, Virgo",
    title = "{GW170817: Observation of Gravitational Waves from a Binary Neutron Star Inspiral}",
    eprint = "1710.05832",
    archivePrefix = "arXiv",
    primaryClass = "gr-qc",
    reportNumber = "LIGO-P170817",
    doi = "10.1103/PhysRevLett.119.161101",
    journal = "Phys. Rev. Lett.",
    volume = "119",
    number = "16",
    pages = "161101",
    year = "2017"
}

@article{LIGOScientific:2018cki,
    author = "Abbott, B. P. and others",
    collaboration = "LIGO Scientific, Virgo",
    title = "{GW170817: Measurements of neutron star radii and equation of state}",
    eprint = "1805.11581",
    archivePrefix = "arXiv",
    primaryClass = "gr-qc",
    reportNumber = "LIGO-P1800115",
    doi = "10.1103/PhysRevLett.121.161101",
    journal = "Phys. Rev. Lett.",
    volume = "121",
    number = "16",
    pages = "161101",
    year = "2018"
}

@article{Miller:2021qha,
    author = "Miller, M. C. and others",
    title = "{The Radius of PSR J0740+6620 from NICER and XMM-Newton Data}",
    eprint = "2105.06979",
    archivePrefix = "arXiv",
    primaryClass = "astro-ph.HE",
    doi = "10.3847/2041-8213/ac089b",
    journal = "Astrophys. J. Lett.",
    volume = "918",
    number = "2",
    pages = "L28",
    year = "2021"
}

@article{Riley:2021pdl,
    author = "Riley, Thomas E. and others",
    title = "{A NICER View of the Massive Pulsar PSR J0740+6620 Informed by Radio Timing and XMM-Newton Spectroscopy}",
    eprint = "2105.06980",
    archivePrefix = "arXiv",
    primaryClass = "astro-ph.HE",
    doi = "10.3847/2041-8213/ac0a81",
    journal = "Astrophys. J. Lett.",
    volume = "918",
    number = "2",
    pages = "L27",
    year = "2021"
}

@article{McLerran:2007qj,
    author = "McLerran, Larry and Pisarski, Robert D.",
    title = "{Phases of cold, dense quarks at large N(c)}",
    eprint = "0706.2191",
    archivePrefix = "arXiv",
    primaryClass = "hep-ph",
    doi = "10.1016/j.nuclphysa.2007.08.013",
    journal = "Nucl. Phys. A",
    volume = "796",
    pages = "83--100",
    year = "2007"
}

@article{McLerran:2018hbz,
    author = "McLerran, Larry and Reddy, Sanjay",
    title = "{Quarkyonic Matter and Neutron Stars}",
    eprint = "1811.12503",
    archivePrefix = "arXiv",
    primaryClass = "nucl-th",
    reportNumber = "INT-PUB-18-060",
    doi = "10.1103/PhysRevLett.122.122701",
    journal = "Phys. Rev. Lett.",
    volume = "122",
    number = "12",
    pages = "122701",
    year = "2019"
}

@article{Jeong:2019lhv,
    author = "Jeong, Kie Sang and McLerran, Larry and Sen, Srimoyee",
    title = "{Dynamically generated momentum space shell structure of quarkyonic matter via an excluded volume model}",
    eprint = "1908.04799",
    archivePrefix = "arXiv",
    primaryClass = "nucl-th",
    reportNumber = "INT-PUB-19-048, APCTP Pre 2019-020",
    doi = "10.1103/PhysRevC.101.035201",
    journal = "Phys. Rev. C",
    volume = "101",
    number = "3",
    pages = "035201",
    year = "2020"
}

@article{Duarte:2020xsp,
    author = "Duarte, Dyana C. and Hernandez-Ortiz, Saul and Jeong, Kie Sang",
    title = "{Excluded-volume model for quarkyonic Matter: Three-flavor baryon-quark Mixture}",
    eprint = "2003.02362",
    archivePrefix = "arXiv",
    primaryClass = "nucl-th",
    reportNumber = "INT-PUB-20-001",
    doi = "10.1103/PhysRevC.102.025203",
    journal = "Phys. Rev. C",
    volume = "102",
    number = "2",
    pages = "025203",
    year = "2020"
}

@article{Duarte:2020kvi,
    author = "Duarte, Dyana C. and Hernandez-Ortiz, Saul and Jeong, Kie Sang",
    title = "{Excluded-volume model for quarkyonic matter. II. Three-flavor shell-like distribution of baryons in phase space}",
    eprint = "2007.08098",
    archivePrefix = "arXiv",
    primaryClass = "nucl-th",
    reportNumber = "INT-PUB-20-028",
    doi = "10.1103/PhysRevC.102.065202",
    journal = "Phys. Rev. C",
    volume = "102",
    number = "6",
    pages = "065202",
    year = "2020"
}

@article{Alford:1997zt,
    author = "Alford, Mark G. and Rajagopal, Krishna and Wilczek, Frank",
    title = "{QCD at finite baryon density: Nucleon droplets and color superconductivity}",
    eprint = "hep-ph/9711395",
    archivePrefix = "arXiv",
    reportNumber = "IASSNS-HEP-97-119, MIT-CTP-2695",
    doi = "10.1016/S0370-2693(98)00051-3",
    journal = "Phys. Lett. B",
    volume = "422",
    pages = "247--256",
    year = "1998"
}

@article{Rapp:1997zu,
    author = {Rapp, R. and Sch{\"a}fer, Thomas and Shuryak, Edward V. and Velkovsky, M.},
    title = "{Diquark Bose condensates in high density matter and instantons}",
    eprint = "hep-ph/9711396",
    archivePrefix = "arXiv",
    reportNumber = "SUNY-NTG-97-30",
    doi = "10.1103/PhysRevLett.81.53",
    journal = "Phys. Rev. Lett.",
    volume = "81",
    pages = "53--56",
    year = "1998"
}

@article{Alford:1998mk,
    author = "Alford, Mark G. and Rajagopal, Krishna and Wilczek, Frank",
    title = "{Color flavor locking and chiral symmetry breaking in high density QCD}",
    eprint = "hep-ph/9804403",
    archivePrefix = "arXiv",
    reportNumber = "IASSNS-HEP-98-29, MIT-CTP-2731",
    doi = "10.1016/S0550-3213(98)00668-3",
    journal = "Nucl. Phys. B",
    volume = "537",
    pages = "443--458",
    year = "1999"
}

@article{Alford:2007xm,
    author = {Alford, Mark G. and Schmitt, Andreas and Rajagopal, Krishna and Sch{\"a}fer, Thomas},
    title = "{Color superconductivity in dense quark matter}",
    eprint = "0709.4635",
    archivePrefix = "arXiv",
    primaryClass = "hep-ph",
    reportNumber = "MIT-CTP-3861",
    doi = "10.1103/RevModPhys.80.1455",
    journal = "Rev. Mod. Phys.",
    volume = "80",
    pages = "1455--1515",
    year = "2008"
}

@article{Jaffe:2004ph,
    author = "Jaffe, R. L.",
    editor = "Kunihiro, Teiji and Onogi, Tetsuya and Abuki, H. and Takahashi, Toru T.",
    title = "{Exotica}",
    eprint = "hep-ph/0409065",
    archivePrefix = "arXiv",
    reportNumber = "MIT-CTP-3538",
    doi = "10.1016/j.physrep.2004.11.005",
    journal = "Phys. Rept.",
    volume = "409",
    pages = "1--45",
    year = "2005"
}

@article{Anselmino:1992vg,
    author = "Anselmino, Mauro and Predazzi, Enrico and Ekelin, Svante and Fredriksson, Sverker and Lichtenberg, D. B.",
    title = "{Diquarks}",
    reportNumber = "TULEA-1992-05, DFTT-7-92, IUHET-215",
    doi = "10.1103/RevModPhys.65.1199",
    journal = "Rev. Mod. Phys.",
    volume = "65",
    pages = "1199--1234",
    year = "1993"
}

@article{Park:2019bsz,
    author = "Park, Aaron and Lee, Su Houng and Inoue, Takashi and Hatsuda, Tetsuo",
    title = "{Baryon{\textendash}baryon interactions at short distances: constituent quark model meets lattice QCD}",
    eprint = "1907.06351",
    archivePrefix = "arXiv",
    primaryClass = "hep-ph",
    reportNumber = "RIKEN-iTHEMS-Report-19",
    doi = "10.1140/epja/s10050-020-00078-z",
    journal = "Eur. Phys. J. A",
    volume = "56",
    number = "3",
    pages = "93",
    year = "2020"
}

@article{Park:2021hqb,
    author = "Park, Aaron and Jeong, Kie Sang and Lee, Su Houng",
    title = "{Case for quarkyoniclike matter from a constituent quark model}",
    eprint = "2105.12948",
    archivePrefix = "arXiv",
    primaryClass = "hep-ph",
    reportNumber = "INT-PUB-21-012",
    doi = "10.1103/PhysRevD.104.094024",
    journal = "Phys. Rev. D",
    volume = "104",
    number = "9",
    pages = "094024",
    year = "2021"
}

@article{Park:2022qew,
    author = "Park, Aaron and Lee, Su Houng",
    title = "{Algebraic approach to a quarkyoniclike configuration and stable diquarks in dense matter}",
    eprint = "2202.05454",
    archivePrefix = "arXiv",
    primaryClass = "nucl-th",
    doi = "10.1103/PhysRevD.105.114034",
    journal = "Phys. Rev. D",
    volume = "105",
    number = "11",
    pages = "114034",
    year = "2022"
}

@article{Park:2022jpf,
    author = "Park, Aaron and Lee, Su Houng",
    title = "{Diquarkyonic matter: Quarks, diquarks, and baryons}",
    eprint = "2212.13030",
    archivePrefix = "arXiv",
    primaryClass = "nucl-th",
    doi = "10.1103/PhysRevD.107.094033",
    journal = "Phys. Rev. D",
    volume = "107",
    number = "9",
    pages = "094033",
    year = "2023"
}

@article{Noh:2024nvp,
    author = "Noh, Sungsik and Park, Aaron and Yun, Hyeongock and Cho, Sungtae and Lee, Su Houng",
    title = "{The inevitable quark three-body force and its implications for exotic states}",
    eprint = "2408.00715",
    archivePrefix = "arXiv",
    primaryClass = "hep-ph",
    doi = "10.1016/j.physletb.2025.139278",
    journal = "Phys. Lett. B",
    volume = "862",
    pages = "139278",
    year = "2025"
}

@article{Baek:2026uup,
    author = "Baek, Jongheon and Park, Aaron and Hiyama, Emiko and Noh, Sungsik and Yun, Hyeongock and Han, Kyong Chol and Lee, Su Houng",
    title = "{Three-body forces in the quark model}",
    eprint = "2608.18949",
    archivePrefix = "arXiv",
    primaryClass = "hep-ph",
    doi = "10.1016/j.ppnp.2026.104277",
    journal = "Prog. Part. Nucl. Phys.",
    pages = "104277",
    year = "2026"
}

@article{Yasutake:2026fjh,
    author = "Yasutake, Nobutoshi and Mukobara, Yuta and Park, Aaron and Lee, Su Houng and Maruyama, Toshiki",
    title = "{Multiquark clustering in neutron-star matter from color-spin molecular dynamics}",
    eprint = "2604.05472",
    archivePrefix = "arXiv",
    primaryClass = "astro-ph.HE",
    doi = "10.1103/154s-hj62",
    journal = "Phys. Rev. D",
    volume = "114",
    number = "4",
    pages = "043050",
    year = "2026"
}

@article{Gal:2019xju,
    author = "Gal, Avraham and Karliner, Marek",
    title = "{A diquark model for the $d^*$ (2380) dibaryon resonance?}",
    eprint = "1905.04652",
    archivePrefix = "arXiv",
    primaryClass = "hep-ph",
    doi = "10.1140/epjc/s10052-019-7024-9",
    journal = "Eur. Phys. J. C",
    volume = "79",
    number = "6",
    pages = "538",
    year = "2019"
}

@article{Otsuka:1978zz,
    author = "Otsuka, T. and Arima, A. and Iachello, F.",
    title = "{Nuclear shell model and interacting bosons}",
    doi = "10.1016/0375-9474(78)90532-8",
    journal = "Nucl. Phys. A",
    volume = "309",
    pages = "1--33",
    year = "1978"
}

@article{Howe:1989,
    author = "Howe, Roger",
    title = "{Remarks on classical invariant theory}",
    journal = "Trans. Amer. Math. Soc.",
    volume = "313",
    pages = "539--570",
    year = "1989"
}

@book{Macdonald:1995,
    author = "Macdonald, I. G.",
    title = "{Symmetric Functions and Hall Polynomials}",
    edition = "2nd",
    publisher = "Oxford University Press",
    address = "Oxford",
    year = "1995"
}

@article{Jaffe:1976yi,
    author = "Jaffe, Robert L.",
    title = "{Perhaps a Stable Dihyperon}",
    reportNumber = "SLAC-PUB-1828",
    doi = "10.1103/PhysRevLett.38.195",
    journal = "Phys. Rev. Lett.",
    volume = "38",
    pages = "195--198",
    year = "1977",
    note = "[Erratum: Phys.Rev.Lett. 38, 617 (1977)]"
}

@article{Abuki:2010jq,
    author = "Abuki, Hiroaki and Baym, Gordon and Hatsuda, Tetsuo and Yamamoto, Naoki",
    title = "{The NJL model of dense three-flavor matter with axial anomaly: the low temperature critical point and BEC-BCS diquark crossover}",
    eprint = "1003.0408",
    archivePrefix = "arXiv",
    primaryClass = "hep-ph",
    reportNumber = "TKYNT-10-02",
    doi = "10.1103/PhysRevD.81.125010",
    journal = "Phys. Rev. D",
    volume = "81",
    pages = "125010",
    year = "2010"
}

@article{Sun:2007fc,
    author = "Sun, Gao-feng and He, Lianyi and Zhuang, Pengfei",
    title = "{BEC-BCS crossover in the Nambu-Jona-Lasinio model of QCD}",
    eprint = "hep-ph/0703159",
    archivePrefix = "arXiv",
    doi = "10.1103/PhysRevD.75.096004",
    journal = "Phys. Rev. D",
    volume = "75",
    pages = "096004",
    year = "2007"
}

@article{Braguta:2016cpw,
    author = "Braguta, V. V. and Ilgenfritz, E. -M. and Kotov, A. Yu. and Molochkov, A. V. and Nikolaev, A. A.",
    title = "{Study of the phase diagram of dense two-color QCD within lattice simulation}",
    eprint = "1605.04090",
    archivePrefix = "arXiv",
    primaryClass = "hep-lat",
    doi = "10.1103/PhysRevD.94.114510",
    journal = "Phys. Rev. D",
    volume = "94",
    number = "11",
    pages = "114510",
    year = "2016"
}

@article{Iida:2022hyy,
    author = "Iida, Kei and Itou, Etsuko",
    title = "{Velocity of sound beyond the high-density relativistic limit from lattice simulation of dense two-color QCD}",
    eprint = "2207.01253",
    archivePrefix = "arXiv",
    primaryClass = "hep-ph",
    reportNumber = "RIKEN-iTHEMS-Report-22",
    doi = "10.1093/ptep/ptac137",
    journal = "PTEP",
    volume = "2022",
    number = "11",
    pages = "111B01",
    year = "2022"
}

@misc{Park:paperII,
    author = "Park, Aaron",
    howpublished = "in preparation"
}

\end{document}